\documentclass[prb,aps,twocolumn,showpacs]{revtex4-1}
\usepackage{amsmath,amssymb,bm}

\usepackage{graphicx}

\begin{document}

\author{Anna Sitek}
 \email{anna.sitek@pwr.wroc.pl}
\author{Pawe{\l} Machnikowski}
 \email{pawel.machnikowski@pwr.wroc.pl}
 \affiliation{Institute of Physics, Wroc{\l}aw University of
Technology, 50-370 Wroc{\l}aw, Poland}

\title{Vacuum-induced coherence in quantum dot systems}

\begin{abstract}
We present a theoretical study of vacuum-induced coherence
 in a pair of vertically stacked semiconductor quantum
dots. The process consists in a coherent excitation transfer from a
single-exciton state localized in one dot to a delocalized state in
which the exciton occupation gets trapped. We study the influence of
the factors characteristic of quantum dot systems (as opposed to
natural atoms): energy 
mismatch, coupling between the single exciton states localized in
different dots, different and non-parallel dipoles due to subband
mixing, as well as coupling to 
phonons. We show that the destructive effect of the energy mismatch
can be overcome by an appropriate interplay of the dipole moments and
coupling between the dots which allows one to observe the trapping effect
even in a structure with technologically realistic energy splitting on
the order of milli-electron-Volts. We also analyze the impact of phonon
dynamics on the occupation trapping and show that phonon effects are
suppressed in a certain range of system parameters. This analysis
shows that the vacuum induced coherence effect and the associated
long-living trapped excitonic population can be achieved in quantum
dots. 
\end{abstract}

\pacs{78.67.Hc,78.47.Cd,03.65.Yz}

\maketitle

\section{Introduction}

Pairs of closely stacked quantum dots (QDs) coupled via inter-band
dipole moments \cite{lovett03b,danckwerts06} (double quantum dots,
DQDs) or by tunneling resulting from carrier wave function overlap and
Coulomb correlations
\cite{bryant93,schliwa01,szafran01,szafran08} (quantum dot molecules, QDMs)
attract much attention due to the richness of their physical
properties which show huge technological promise for nanoelectronics,
spintronics and quantum information processing applications. The
unique features of these systems, as compared to individual QDs, can
be used as the basis 
for long-time storage of quantum information \cite{pazy01b} and
conditional optical control of carrier states \cite{unold05} which
pave the way to an implementation of a two-qubit quantum gate
\cite{biolatti00}. Double dot structures enable also coherent optical spin
control and entangling \cite{troiani03,nazir04,gauger08b} or may act as
sources of entangled photons \cite{gywat02}. It has been shown that
the exciton spectrum of a QDM can be used to define an excitonic qubit
with an extended life time \cite{rolon10}, that information can be
written on the spin state of the dopant Mn ion located in one of the
dots forming a DQD \cite{goryca09}, and that a photon emitted by a
nearby quantum point contact may induce carrier transfer in a DQD
system \cite{gustavsson07}. The richness and complexity of the
physical properties of these systems have been manifested in many
optical experiments \cite{borri03,bardot05,gerardot05}.

The spacing between the dots forming intentionally manufactured DQDs
and QDMs is typically on the order of nanometers which is two to three
orders of magnitude smaller than the wavelength of a resonantly
coupled photon. Under such conditions, in atomic samples the effect of
superradiant emission is observed \cite{dicke54}. A similar effect of
coupling to common 
radiative modes appears also in the emission from QD systems. 
Here 
also superradiance-like phenomena occur in the evolution of the
exciton occupation and polarization.
\cite{scheibner07,averkiev09,yukalov10,kozub12}.
Apart from the modified evolution for spontaneous emission, collective
coupling to the radiation field can lead to other physical effects,
one of which is the vacuum induced coherence (VIC)
\cite{agarwal74}. This effect consists in spontaneous, partial
coherent excitation transfer from an initially occupied QD to
initially empty one which results in exciton occupation trapping in a
decoherence-resistant state. Like the superradiance phenomena, the
VIC effect can be expected to occur also in QD
systems.

On the other hand, although natural atoms and QDs have many properties
in common, the 
characteristic feature of the latter is the inhomogeneity of
transition energy which, for technologically realistic structures, is
on the order of milli-electron-Volts. 
Moreover, small spatial separation of the dots leads to coupling
between them \cite{bayer01}. In our previous works we studied the
impact of the transition energy mismatch and the coupling between the
two emitters on the stability of collective effects
\cite{sitek11,sitek07a} and their role in the linear \cite{sitek09b}
and nonlinear optical response of the DQD systems
\cite{sitek09a}. We showed that, in the absence of coupling the
between the QDs, the appearance of an optically inactive subradiant
state and a rapidly decaying superradiant one are extremely sensitive
to the fundamental transition energy mismatch and already for a
mismatch on the order of micro-electron-Volts the collective character of
the evolution is replaced by oscillation around the average
exponential decay.  The destructive effect of the system inhomogeneity
may be, to some extent, overcome by sufficiently strong coupling
between the dots which rebuilds collective behavior even in structures
with technologically realistic values of energy mismatch
\cite{sitek07a,averkiev09}. We have also 
pointed out that phonon-induced dynamics can slow down the decay of a
superradiant state or speed up the emission from the superradiant one
\cite{machnikowski09c}.
Based on these previous investigations of superradiance phenomena in
QDs, one can expect that the VIC effect should, in principle, also be
observable in these systems,
at least in the presence of sufficiently strong coupling
\cite{sitek11}. However, to our knowledge, its stability against 
various inhomogeneities and perturbations typical for the solid state
environmnet (in particular non-parallel orientation of the interband
dipoles in the two dots as well as
phonon perturbations) have not been studied.

In this paper, we study the necessary conditions for the VIC effect to
appear in a system of two vertically stacked semiconductor QDs. We
investigate the role of the fundamental transition energy mismatch,
coupling between the dots, and phonon-induced kinetics in
this process. We also pay particular attention to the difference of
the magnitude 
of the interband dipole moments as well as to their non-parallel
alignment (due 
to subband mixing). We show that in spite of all these inhomogeneities
that are typical to QD structures (as opposed to natural atoms) the
vacuum-induced coherence can be almost fully stabilized in realistic
pairs of non-identical QDs in a certain range of parameters. In
particular, different interband dipole moments for the two dots lead
to the appearance of a state which is perfectly immune to radiative
decay for a particular 
choice of the system parameters.

The paper is organized as follows. In Sec.~\ref{sec:system}, we describe the
system under investigation and define its model.  In
Sec.~\ref{sec:evolution}, the method for describing the evolution is 
described. Section \ref{sec:results} contains the discussion of the
results. Concluding remarks are contained in Section \ref{sec:concl}.

\section{The system}
\label{sec:system}

The investigated system is composed of two vertically stacked semiconductor QDs interacting with quantum electromagnetic field 
and lattice vibrations. We restrict the discussion to the ground-level transitions with fixed polarization and spin orientations. 
We take into consideration only `spatially direct' states in which electron-hole pairs reside in the same QD. Due to the strong 
Coulomb coupling these states have a much lower energy than the `dissociated' states (the external electric fields which would 
change this picture \cite{szafran05,szafran08} are not considered in our discussion). In this manner, the DQD or QDM may be 
described as a four-level system, with the state $|00\rangle$ denoting empty dots, states $|10\rangle$ and $|01\rangle$ 
representing single-exciton states with electron-hole pairs residing in the lower or higher QD, and the $|11\rangle$ corresponding 
to the biexciton state, that is, to both QDs occupied with an exciton.

As it is currently impossible to produce on demand pairs of QDs with
identical fundamental transition energies, we assume that the  
exciton transition energies for the two dots are different, 
\begin{displaymath}
E_1=E+\Delta \quad \mathrm{and} \quad E_2=E-\Delta,
\end{displaymath}
where $E$ is the average transition energy and $\Delta$ is the energy mismatch.

As in Ref.~\onlinecite{sitek09b}, we describe the evolution in the
`rotating basis' defined by the unitary transformation  
\begin{displaymath}
U=e^{-i\left[ E\left(|10\rangle\!\langle 10| + |01\rangle\!\langle 01|
+ 2|11\rangle\!\langle 11|\right)
+H_{\mathrm{rad}}+H_{\mathrm{ph}}\right]t/\hbar},
\end{displaymath}
where $H_{\mathrm{rad}}$ and $H_{\mathrm{ph}}$ are the standard free
photon and phonon Hamiltonians, respectively. The Hamiltonian of the system is then
\begin{equation}
\label{hamiltonianH}
H=H_{\mathrm{DQD}}+H_{\mathrm{DQD-ph}}+H_{\mathrm{DQD-rad}}.
\end{equation}
The first term describes exciton states in a DQD structure.
\begin{eqnarray}
H_{\mathrm{DQD}}&=&\Delta\big(|10\rangle\!\langle
10|-|01\rangle\!\langle 01|\big) +V_{\mathrm{B}}|11\rangle\!\langle 11|
\nonumber \\
&+&V\big(|10\rangle\!\langle 01|+|01\rangle\!\langle 10|\big),
\label{hamiltonianDQD}
\end{eqnarray}
where $V_{\mathrm{B}}$ is a biexciton shift due to the interaction of
static dipole moments and $V$ is the amplitude of the coupling  
between the single-exciton states of the dots.

The QDs are separated by a distance on the order of a few nm which is
much smaller than the relevant wave length of the electromagnetic  
field with which the dots interact. This allows us to neglect the
space dependence of the electromagnetic field and describe the  
coupling of excitons to the photon modes in the Dicke limit
\cite{dicke54}. The relevant Hamiltonian in the dipole  and rotating  
wave approximations is then
\begin{displaymath}
\label{hamiltonianDQDrad}
H_{\mathrm{DQD-rad}}=\sum_{\alpha=1}^{2}\sum_{\bm{k}\lambda}
\sigma^{\alpha}_{-}g^{\alpha}_{\bm{k}\lambda}
e^{-i\left(\frac{E}{\hbar}-\omega_{\bm{k}}\right)t}b^{\dagger}_{\bm{k},\lambda} 
+\mathrm{H.c}.
\end{displaymath}
where $\sigma^{\alpha}_{+}=\left(\sigma^{\alpha}_{-}\right)^\dagger$
are the creation and annihilation operators for the exciton in  
the $\alpha$th QD, $b^{\dagger}_{\bm{k}}$  is the creation operator of
the photon mode with the wave vector $\bm{k}$, and 
\begin{displaymath}
g^{\alpha}_{\bm{k}\lambda}=i\bm{d}_{\alpha}\cdotp\!\hat{e}_{\lambda}(\bm{k})
\sqrt{\frac{\hbar\omega_{\bm{k}}}{2\epsilon_{0}\epsilon_{r}v}}
\end{displaymath} 
is a coupling constant for the $\alpha$th QD, where $\bm{d}_{\alpha}$
is the inter-band dipole moment for the $\alpha$th  
QD, $\hat{e}_{\lambda}(\bm{k})$ is the unit polarization vector of the
photon mode with polarization $\lambda$,  
 $\omega_{\bm{k}}$ is the corresponding frequency, $\epsilon_0$ is the
 vacuum dielectric constant, $\epsilon_r$  
is the relative dielectric constant of the semiconductor and $v$ is
the normalization volume. We investigate only  
wide-gap semiconductors with fundamental transition energies on the
order of 1 eV for which zero-temperature approximation  
for the electromagnetic modes may be used at any reasonable
temperature.

Interaction of the carriers confined in the two dots with phonon modes is modeled by the Hamiltonian
\begin{displaymath}
H_{\mathrm{DQD-ph}}=\sum\limits_{\alpha=1, 2}\sigma^{\alpha}_{+}\sigma^{\alpha}_{-}
\sum_{\bm{q}}f^{(\alpha)}_{\bm{q}} (c_{\bm{q}}+c_{-\bm{q}}^{\dagger}),
\end{displaymath}
where $c_{\bm{q}}^{\dagger}$ and $c_{\bm{q}}$ are creation and
annihilation operators of the phonon mode with the wave vector  
$\bm{q}$ and $f^{(1,2)}_{\bm{q}}$ are the system reservoir coupling
constants for the first and second QD, respectively. We model  
the electron and hole wave functions by identical Gaussians with 
extensions $l$ in the $xy$ plane and $l_{z}$ along the growth direction,
\begin{displaymath}
\Psi(\bm{r})\sim\exp\left(-\frac{1}{2}\frac{x^{2}+y^{2}}{l^{2}}-\frac{1}{2}\frac{z^{2}}{l_{z}^{2}}\right).
\end{displaymath}   
For such wave functions and for the deformation potential couplings
between the confined carriers and longitudinal phonon modes,  
the coupling constants have the form\cite{machnikowski09c}
\begin{displaymath}
f^{(1,2)}_{\bm{q}}=f_{\bm{q}}\exp\left[\pm \frac{iq_{z}D}{2}\right],
\end{displaymath}
where $D$ is the distance between the dots and
\begin{displaymath}
f_{\bm{q}}=(\sigma_{\mathrm{e}}-\sigma_{\mathrm{h}})
\sqrt{\frac{\hbar q}{2\rho v c_{\mathrm{l}}}}
\exp\left[-\frac{l^{2}_{z}q^{2}_{z}+l^{2}q^{2}_{\bot}}{4}\right].
\end{displaymath}
Here $\sigma_{\mathrm{e/h}}$ are deformation potential constants for
electrons/holes, $\rho$ is the crystal density, $c_{\mathrm{l}}$ is
the speed of longitudinal 
sound (linear phonon dispersion is assumed) and $q_{\bot,z}$ are
momentum components in the $xy$   
plane and along the $z$ axis.

\section{The evolution}
\label{sec:evolution}

Analytical formulas describing the evolution of a pair of QDs are
available for uncoupled systems ($V=0$) interacting only with phonon
modes \cite{roszak06a} and, in the Markov limit, if only radiative
decay is included \cite{sitek07a}. In this paper we use the previously
proposed method \cite{machnikowski09b} that allows us to deal with the
simultaneous action of both the phonon and photon surroundings. Our
approach is based on the equation of motion for the reduced density
matrix of the carrier subsystem in the interaction picture,
\begin{equation}
\label{evolution}
\dot{\rho}=\mathcal{L}_{\mathrm{rad}}[\rho]+\mathcal{L}_{\mathrm{ph}}[\rho].
\end{equation}  
Here the first term accounts for the effects induced by the radiative environment, which is described in the Markov limit by the 
Lindblad dissipator
\begin{eqnarray}
\lefteqn{\mathcal{L}_{\mathrm{rad}}[\rho]=} \nonumber \\
&& \sum_{\alpha\beta=1}^{2}\Gamma_{\alpha\beta}
\left[\sigma^{\alpha}_{-}(t)\rho\sigma^{\beta}_{+}(t)
-\frac{1}{2}\left\{\sigma^{\beta}_{+}(t)\sigma^{\alpha}_{-}(t),\rho\right\}\right],
\label{superLindblad}
\end{eqnarray}
where
\begin{displaymath}
\sigma^{\alpha}_{-}(t)=\left[\sigma^{\alpha}_{+}(t)\right]^{\dagger}=
\exp\left[\frac{iH_{\mathrm{DQD}}t}{\hbar}\right]\sigma^{\alpha}_{-}
\exp\left[-\frac{iH_{\mathrm{DQD}}t}{\hbar}\right] 
\end{displaymath}
and 
\begin{equation}
\label{Gamma}
\Gamma_{\alpha\beta}=\Gamma^{*}_{\beta\alpha}=
\frac{E^{3}}{3\pi\epsilon_{0} \epsilon_{r}\hbar^{4}}
\bm{d}_{\alpha}\!\cdot\!\bm{d}^{*}_{\beta}.
\end{equation}
If we assume that the spontaneous decay rates for the two QDs are
$\Gamma_{11}$ and $\Gamma_{22}$ then it follows directly from  
Eq. (\ref{Gamma}) that 
\begin{displaymath}
\Gamma_{12}=\Gamma_{21}^{*}
=\sqrt{\Gamma_{11}\Gamma_{22}}\hat{\bm{d}}_{1}\!\cdot\!\hat{\bm{d}}^{*}_{2},
\end{displaymath}
where $\hat{\bm{d}}_{\alpha}=\bm{d}_{\alpha}/d_{\alpha}$.
By redefining the relative phase of the exciton states in the two dots
one can assume without any loss of generality that
$\Gamma_{12}$ and $\Gamma_{21}$ are real.

The second term in Eq.~(\ref{evolution}) accounts for the effects due
to interaction with phonon surrounding, allowing for non-Markovian
dynamics. To describe these effects we use the time-convolutionless
equation 
\begin{eqnarray*}
\lefteqn{\mathcal{L}_{\mathrm{ph}}[\rho]=}\\
&  &  -\int\limits^{t}_{0}d\tau\mathrm{Tr_{ph}}\bigl[H_{\mathrm{DQD-ph}}(t),
\left[H_{\mathrm{DQD-ph}}(\tau),\rho(t)\otimes\rho_{\mathrm{ph}}\right]\bigr],
\end{eqnarray*}
where
\begin{displaymath}
H_{\mathrm{DQD-ph}}(t)=
\exp\left[\frac{iH_{\mathrm{DQD}}t}{\hbar}\right]H_{\mathrm{DQD-ph}}\exp\left[-\frac{iH_{\mathrm{DQD}}t}{\hbar}\right],
\end{displaymath}
$\rho_{\mathrm{ph}}$ is the phonon density matrix at thermal equilibrium, and $\mathrm{Tr_{ph}}$ denotes partial trace with 
respect to the phonon degrees of freedom.

%
\vspace{0.2 cm}
\begin{center} 
\begin{table}[tb]
\begin{tabular}{l c c}\hline\hline
Parameter & Symbol & Value\\ \hline
Difference of deformation potential  & &  \\
constants for electrons and holes & $\sigma_{e}-\sigma_{h}$ & $9$ meV \\
Crystal density & $\varrho$ & $5350$ kg/m$^{3}$ \\
Speed of longitudinal sound & $c_{\mathrm{l}}$ & $5150$ m/s \\
Carrier localization extensions & & \\
in the $xy$ plane & $l$ & $4.5$ nm \\ 
Carrier localization extensions & & \\
in the growth direction & $l_{z}$ & $1$ nm \\ 
Spatial separation of the dots & $D$ & 8 nm \\
\hline\hline 
\label{tabela}
\end{tabular}
\caption{Parameters used in numerical simulations. The values
  correspond to a self-assembled InAs/GaAs system.}  
\end{table}
\end{center}

\section{Results and discussion}
\label{sec:results}

Below, we present our results of simulations of the VIC process in
pairs of QDs. In all the analyzed cases, we assume that the system is  
prepared initially in a localized state $|10\rangle$. In
Sec. ~ref{subsec:identical} we explain the effect in a system of
identical, uncoupled dots. Then, in Sec.~\ref{subsec:dipole}, we
analyze the role of the relative magnitude and orientation
dipole moments and in Sec.~\ref{subsec:DV} the effect of the   
energy mismatch and coupling between the dots. In
Sec.~\ref{subsec:all} we analyze the interplay of all the parameters that  
distiguish QDs from natural atoms in the evolution of QDs interacting only
with radiation reservoir. The phonon impact on the  
VIC is discussed in Sec. \ref{subsec:fonony}.

\subsection{Identical QDs}
\label{subsec:identical}

\begin{figure}[t]
\centering
\includegraphics[scale=0.7]{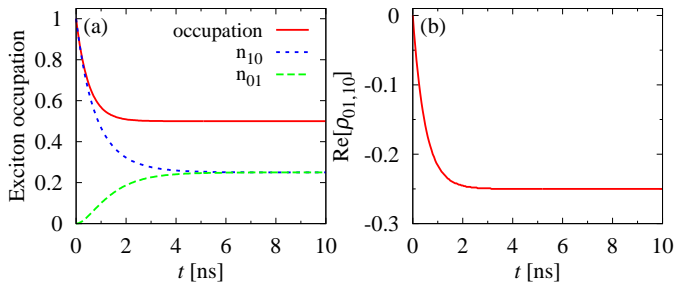}
\caption{(Color online) Vacuum-induced coherence in a system of two identical
  ($\Delta=0$, $\Gamma_{11}=\Gamma_{22}=1$ ns$^{-1}$)  
two-level systems. (a) Exciton occupation of the system and
occupations of the states $|10\rangle$ ($n_{10}$) and  
$|01\rangle$ ($n_{01}$). (b) The off-diagonal density matrix element
$\rho_{01,10}$.} 
\label{fig:VIC_Delta=0}
\end{figure}

Collective coupling of two identical QDs (systems with identical
fundamental transition energies and parallel dipole moments of equal
magnitudes) to the quantum electromagnetic vacuum leads to the
appearance of a short-living (bright) superradiant state, $|+\rangle =
\left(|10\rangle+|01\rangle\right)/\sqrt{2}$, and an optically
inactive (dark) subradiant state $|-\rangle =
\left(|10\rangle-|01\rangle\right)/\sqrt{2}$. The initial state of the
analyzed system is a localized single-exciton state ($|10\rangle$ or
$|01\rangle$) which can appear naturally, e.g., as an effect of incoherent
trapping or controlled tunnel injection of carriers in an injection
structure similar to that studied in Ref.~\onlinecite{rudno06}. Such a state
of a system of two identical and uncoupled QDs may be expressed as an
equal combination of the sub- and superradiant states, $|-\rangle$ and
$|+\rangle$,
%
%
%
%
%
%
\begin{equation}
|10\rangle = \frac{1}{\sqrt{2}}\left(|+\rangle+|-\rangle\right),\quad
|01\rangle = \frac{1}{\sqrt{2}}\left(|+\rangle-|-\rangle\right).
\end{equation}
Coupling to the electromagnetic reservoir induces emission from the
bright state and, consequently, decay of a half of the  
initial excitation, while the other half is unaffected (trapped) due to the
stability of the subradiant state for a system of two identical  
two-level systems \cite{sitek07a}. This is observed as a spontaneous
and coherent excitation transfer from initially occupied dot  
to the initially empty one until occupation of both QDs stabilizes at
the same level (Fig.~~\ref{fig:VIC_Delta=0}a). While this   
process is taking place, coherence builds up spontaneously in the
system (Fig.~\ref{fig:VIC_Delta=0}b). As a consequence, the pair  
of identical QDs is trapped in a delocalized and decoherence-resistant
state with the exciton occupation number equal $0.5$ and  
real off-diagonal matrix element equal $-0.25$. This effect is
referred to as VIC.

\subsection{The role of dipole moments}
\label{subsec:dipole}

\begin{figure}[t]
\centering
\includegraphics[scale=0.7]{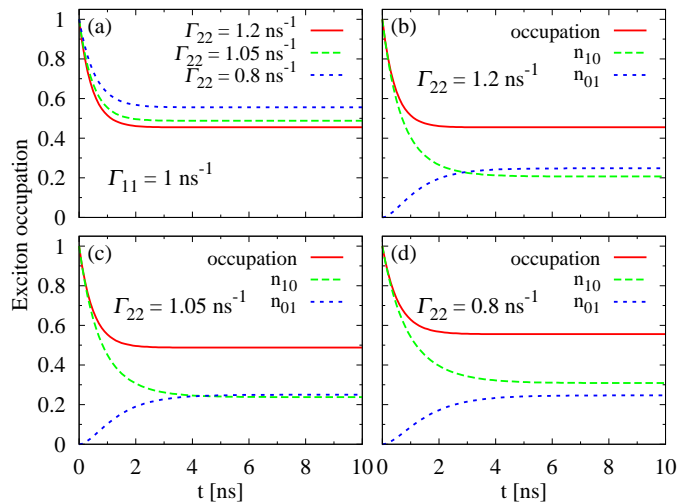}
\caption{(Color online) (a) Exciton occupation (red solid line) for a
  pair of decoupled ($V=0$) QDs with 
  identical transition energies ($\Delta=0$) but with  
non-identical values of the spontaneous recombination rates
($\Gamma_{11}\neq\Gamma_{22}$). 
(b-d) The total exciton occupation (red solid line) and the
occupations of the individual dots (green and blue dashed lines) for
the three cases shown in (a). 
The value of $\Gamma_{11}=1$ ns$^{-1}$ 
is the same for all the graphs in this figure.}
\label{fig:Delta=0_rozne_dipole}
\end{figure}

Vertically stacked semiconductor QDs differ slightly in size and
shape. If the system was formed in a self-assembled two-layer  
process then the upper QD is usually bigger than the lower one
\cite{heitz98}. The geometry and the structure of the QDs is  
reflected in the carrier wave functions and thus in the interband
dipole moments. According to Eq.~(\ref{Gamma}), this leads to
different   spontaneous decay rates for the two QDs forming the
investigated system. The dipoles corresponding to the upper and lower
QD  may differ in amplitude and, if the hole states in the two
structures have different light-hole admixtures, also in orientation
(see Appendix).  

The equation of motion describing the evolution of a system of two
energetically identical ($\Delta=0$) and uncoupled ($V=0$) QDs  
interacting only with the radiation reservoir is described by
Eq.~\eqref{evolution} with $\mathcal{L}_{\mathrm{ph}}=0$ and
$\mathcal{L}_{\mathrm{rad}}$ given by 
 Eq. (\ref{superLindblad}) with
 $\sigma_{\pm}^{\alpha}(t)=\sigma_{\pm}^{\alpha}$. 
It is known that
in three-level open systems of this kind, non-radiating superposition
states occur under certain conditions
\cite{cardimona82,hegerfeldt92,carvalho01,santos12}.
In our case, a non-trivial (different than the ground state
$|00\rangle\!\langle 00|$)  
stationary solution to the open system evolution equation, corresponding
to the spontaneously formed, stable,
delocalized dark state discussed above,
exists only for $\Gamma_{12}=\sqrt{\Gamma_{1}\Gamma_{2}}$, i.e., for
parallel dipole moments [Eq. (\ref{Gamma})]
(such that $\hat{\bm{d}}_{1}=\hat{\bm{d}}_{2}$ up to a phase factor). 
For dipoles that are parallel but have different amplitudes, the
evolution is similar to the case of identical systems, i.e.,  
coupling to photon surrounding leads to excitation transfer and
occupation trapping. As can be seen in  
Fig.~\ref{fig:Delta=0_rozne_dipole}, the fraction of trapped exciton
occupation depends on the values of the single  
dot decay rates and stabilizes at the level
$\Gamma_{11}/\left(\Gamma_{11}+\Gamma_{22}\right)$.

As in the case of identical QDs, the suppression of the exciton decay
is due to the existence of a dark state which, for  
$\Gamma_{11}\neq\Gamma_{22}$, is not strictly anti-symmetric,
\begin{equation}
\label{darkG11G22}
|\mathrm{dark}\rangle =
 \frac{\sqrt{\Gamma_{11}}|10\rangle-\sqrt{\Gamma_{22}}|01\rangle}{
\sqrt{\Gamma_{11}+\Gamma_{22}}}.
\end{equation}
Due to unequal contribution from the single-exciton states to the dark
state (\ref{darkG11G22}) the final occupation of the dots  
is non-symmetric, either, and equals
\begin{displaymath}
 n_{10}=\left(\frac{\Gamma_{11}}{\Gamma_{11}+\Gamma_{22}}\right)^{2} \quad \mathrm{and} \quad 
n_{01}= \frac{\Gamma_{11}\Gamma_{22}}{\left(\Gamma_{11}+\Gamma_{22}\right)^{2}},
\end{displaymath}
respectively. Different dipole moments allow one to achieve different
final situations. In the trapped state, the occupation of the
initially excited dot may be lower
[Figs.~\ref{fig:Delta=0_rozne_dipole}(b) and
\ref{fig:Delta=0_rozne_dipole}(c)]  or higher 
[Fig.~\ref{fig:Delta=0_rozne_dipole}(d)] than that of the
initially empty dot.

\begin{figure}[t]
\centering
\includegraphics[scale=0.7]{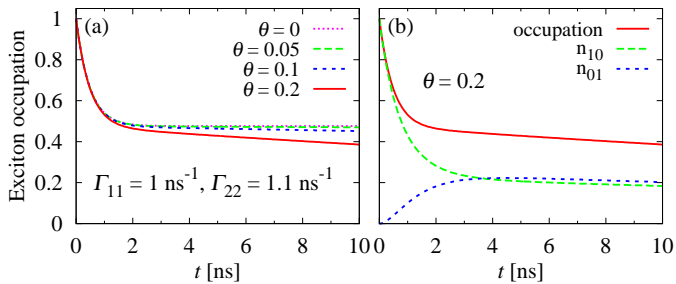}
\caption{Exciton occupation of a pair of uncoupled ($V=0$) QDs with
  identical transition energies ($\Delta=0$) and different  
spontaneous recombination rates ($\Gamma_{11}\neq\Gamma_{22}$) for a
few values of the angle $\theta$ between the dipole  
moments. The values of $\Gamma_{11}$ and $\Gamma_{22}$ shown in
Fig.~(a) are valid for both figures.} 
\label{fig:teta_rozne_dipole}
\end{figure}

Another factor that influences the interband dipole moments is the
light-hole admixture. If this admixture is different for the  
two dots then the dipole moments corresponding to the QDs forming the
QDM or DQD become non-parallel (see Appendix). This means that  
the off-diagonal decay rate
$\Gamma_{12}=\sqrt{\Gamma_{11}\Gamma_{22}}
\left(1-\theta^{2}/2\right)\neq\sqrt{\Gamma_{11}\Gamma_{22}}$,  
where $\theta$ is the angle between the interband dipole moments of
the two dots 
[see Eq.~\eqref{dipole2}]. Consequently, the stationary solution to
Eq.~\eqref{superLindblad} 
does not exist and quenching of the final exciton occupation is
observed. The values of $\theta$ are determined by the light hole
admixture and typically are on the order  
of $0.01$ \cite{gawarecki12a}. However, as can be seen in
Fig.~\ref{fig:teta_rozne_dipole}, even for much larger values of the
angle between the  
dipoles, quenching of the occupation is weak. Although the system
looses its coherence at long times, the character of the  
evolution remains the same as in the case of parallel dipoles on time
scales much longer than the nominal 
exciton lifetime, i.e., initially  
excitation transfer between the dots takes place until the occupation
rate $n_{10}/n_{01}$ close to that defined by the decay  
rates $\Gamma_{11}$ and $\Gamma_{22}$ is reached and, then the impact of of hole subband mixing is manifested as a slow and 
equal decay of occupations [Fig.~\ref{fig:teta_rozne_dipole}(b)].      

\subsection{The role of the energy mismatch and 
coupling between the dots}
\label{subsec:DV}

\begin{figure}[t]
\centering
\includegraphics[scale=0.7]{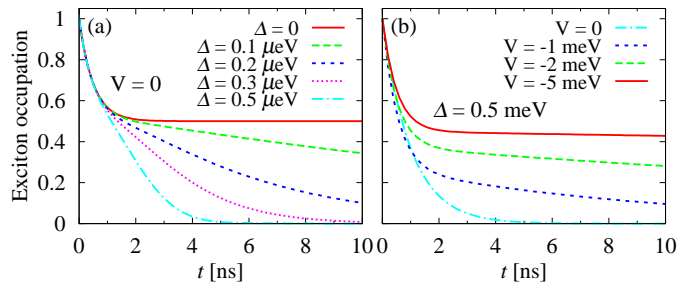}
\caption{(a) The impact of the energy mismatch on exciton occupation
  of a pair of uncoupled QDs.  
(b) Exciton occupation of a technologically realistic QDM for a few
values of coupling between the emitters  
($\Gamma_{11}=\Gamma_{22}=1$ ns$^{-1}$).}
\label{fig:VIC_D_V}
\end{figure}

In spite of rapid technological progress, manufacturing of DQDs or QDMs with
identical fundamental transition energies is still not feasible. As
can be seen in  Fig.~\ref{fig:VIC_D_V}(a), for the realistic case of
non-zero energy mismatch, the effect of trapping the system in an
occupied  
and optically inactive state is destroyed. Already for the energy
splitting on the order of tens of $\mu$eV quenching of the final
occupation is observed  (the relevant energy 
scale is the transition energy line width
$\hbar\Gamma=1.7\,\mu$eV). QDs constituting such an   
inhomogeneous double dot systems interact with disjoint energy ranges
of the electromagnetic field which destroys the collective character  
of coupling to photon modes. The energy mismatch of the dots slows the
decay of the superradiant state and induces emission  
from the subradiant one \cite{sitek07a}. Due to the lack of a stable
exciton state in which the system might be trapped the  
VIC effect in inhomogeneous QDMs is destroyed.

Initially, the character of the evolution of an exciton occupation of
the inhomogeneous pairs of QDs does not differ considerably from the
corresponding case of a pair of identical dots. Until
$t\sim\hbar/(2\Delta)$, the coupling to photon modes maintains its
collective character and excitation transfer from the initially
occupied QD to the initially empty one takes place. Later, due to the
emission from the subradiant as well as from the superradiant state,
occupations of both dots decay \cite{sitek11}.

If the distance between the QDs is sufficiently small coupling between
the systems (F{\"o}rster or tunneling) becomes effective and affects
the evolution of carriers confined in the structure. Since sub- and
superradiant states are eigenstates of the coupling part of the
Hamiltonian [Eg.(\ref{hamiltonianDQD})] separated by the energy $2V$,
sufficiently strong interaction between the dots rebuilds the
collective character of the interaction even in structures with
technologically realistic energy mismatches on the order of 1 meV
\cite{sitek07a,gerardot05}. This also enables the VIC effect in
inhomogeneous DQDs to be rebuilt. As can be seen in
Fig.~\ref{fig:VIC_D_V}(b), for $V\gg\Delta$, a transition from a
localized initial single-exciton state to a nearly stable state is
possible for a system with a technologically achievable energy
mismatch. Although full stabilization of the VIC effect is
impossible for non-identical dots and quenching of the exciton
occupation always takes place
sufficiently strong coupling between the dots considerably reduces the
decay of the trapped state. For a 
weaker coupling between the dots, the exciton occupation decay is
faster but still reduced 
compared to the system of uncoupled QDs.  In contrast to
identical dots, the two localized states have different contributions
from the from sub- and superradiant states. This results in a
different degree of trapping depending on the choice of
the initially occupied state \cite{sitek11}.

\subsection{Interplay of energy mismatch, coupling between the dots and non-uniform dipoles}
\label{subsec:all}

\begin{figure}[t]
\centering
\includegraphics[scale=0.7]{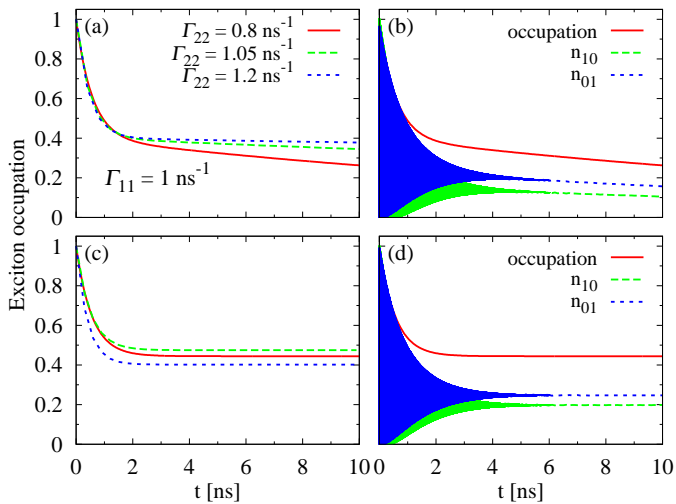}
\caption{Exciton occupation of a pair of technologically realistic
  QDs. 
(a) Total exciton occupation for $\Delta=1$~meV, $V=-5$~meV,
$\Gamma_{11}=1$~ns$^{-1}$, and for a few values of $\Gamma_{22}$. (b)
The total occupation and the occupations of the individual dots for
the last case shown in (a). (c) Exciton occupation for three sets of
parameters stabilizing the occupation trapping. 
Red solid line: $\Delta=0.5$~ meV, $V=-4.47$~meV,  
$\Gamma_{11}=0.8$~ns$^{-1}$,  $\Gamma_{22}=1$~ns$^{-1}$. 
Green dashed line: $\Delta=0.5$~meV, $V=-9.99$~meV, 
$\Gamma_{11}=0.95$~ns$^{-1}$, $\Gamma_{22}=1.05$~ns$^{-1}$. 
Blue dotted line: $\Delta=1$~meV, $V=-5$~meV,
$\Gamma_{11}=1$~ns$^{-1}$, $\Gamma_{22}=1.49$~ns$^{-1}$.
(d) The total occupation and the occupations of the individual dots
corresponding to the red line in (c).}
\label{fig:realistic}
\end{figure}

For technologically realistic DQDs and QDMs, the fundamental energy
mismatch of the two QDs forming the system is on the order of  
milli-electron-Volts, the dipole moments differ slightly between the
dots and the systems are coupled with one another via  
F{\"o}rster or tunneling coupling. The evolution of such systems, for
parallel dipole moments, is shown in  
Fig.~\ref{fig:realistic}. In Fig.~\ref{fig:realistic}(a), the dynamics
of a system strongly stabilized by a coupling between the  
dots is presented for a few values of the decay rates. As can be seen,
the rate of occupation decay depends on the choice of 
the two single-dot decay rate parameters. 
In such a system, the coherent excitation transfer between the dots
takes place on time scales similar to the previously discussed
cases. However, the initial state is now a superposition of
non-degenerate system eigenstates, which leads to  fast oscillations
of the exciton occupation between the two dots [Fig.~\ref{fig:realistic}(b)]. 

In the present general case, the spontaneously formed coherence and
the trapped occupation can again be fully stabilized by an appropriate
choice of parameters in a system with parallel dipoles. 
The single-exciton eigenstates of the system Hamiltonian
(\ref{hamiltonianH}) are 
%
%
%
%
%
%
%
%
%
%
\begin{subequations}
\begin{eqnarray}
\label{wlasne1}
 |\Psi_{1}\rangle &=& \cos\!\left(\frac{\phi}{2}\right)|10\rangle+\sin\!\left(\frac{\phi}{2}\right)|01\rangle,\\
\label{wlasne2}
 |\Psi_{2}\rangle &=& -\sin\!\left(\frac{\phi}{2}\right)|10\rangle+\cos\!\left(\frac{\phi}{2}\right)|01\rangle,
\end{eqnarray}
\end{subequations}
where $\mathrm{tg}(\phi)=V/\Delta$ and $\pi/2 \leqslant\phi<\pi/2$.
The corresponding eigenenergies will be denoted by $E_{1},E_{2}$. 
If
the coupling between the single exciton states satisfies the relation
\begin{equation}
\label{VVVV}
  V=-\frac{2\Delta\sqrt{\Gamma_{11}\Gamma_{22}}}{
|\Gamma_{22}-\Gamma_{11}|}
\end{equation}
then one of the eigenstates [Eqs. (\ref{wlasne1}) and (\ref{wlasne2})]
corresponds to the dark state (\ref{darkG11G22}). In such case, the
evolution of a  
realistic system also leads to occupation trapping in a
decoherence-resistant state [Fig.~\ref{fig:realistic}(c)] at the  
occupation $\Gamma_{11}/(\Gamma_{11}+\Gamma_{22})$. As can be seen
from Eq.~ (\ref{VVVV}), non-equal values  
of the spontaneous decay rate for the two dots open a possibility to stabilize the evolution of the system for wide range of 
parameters. As for a system of identical dots, the occupations of
single dots also stabilizes, but again initial fast oscillations 
due to contribution from two energy eigenstates are observed
[Fig.~\ref{fig:realistic}(d)].

\begin{figure}[t]
\centering
\includegraphics[scale=0.7]{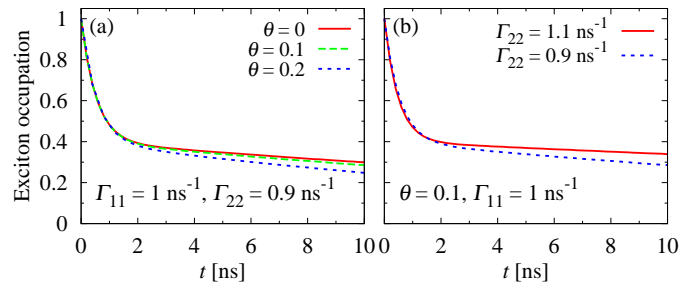}
\caption{Exciton occupation of a pair of non-identical ($\Delta=1$ meV, $\Gamma_{11}\neq\Gamma_{22}$) and coupled ($V=-5$ meV) QDs for 
a few values of $\theta$ (a) and for two values of $\Gamma_{22}$ (b).}
\label{fig:realisticQDs_teta}
\end{figure}

For the non-parallel dipole moments, i.e., for $\theta\neq 0$, the
effect of occupation trapping is in principle destroyed but
for strongly coupled dots, the 
decay is very weak. For the interesting time scale on the order of ns
the impact of non-parallel dipoles becomes visible only for the values of  
$\theta$ that exceed the realistic ones by an order of magnitude
\cite{gawarecki12a} (Fig.~\ref{fig:realisticQDs_teta}).

\subsection{The role of lattice dynamics}
\label{subsec:fonony}

\begin{figure}[tb]
\begin{center}
\includegraphics[width=60mm]{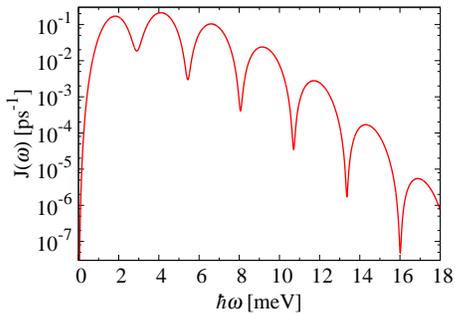}
\end{center}
\caption{\label{fig:spdens}(Color online) The phonon spectral density
  for $D=8$~nm.}
\end{figure}

Spontaneous emission from a system of QDs is affected by phonon
dynamics \cite{karwat11,machnikowski09c}. Coupling between the QDs and
lattice  vibrations induces   
excitation transfer between the two single exciton eigenstates
[Eq.~\eqref{wlasne1}and \eqref{wlasne2}].
For a pair of identical QDs ($\phi=\pm\pi/2$), these eigenstates
exactly coincide with the bright (superradiant) and dark  (subradiant)
states $|+\rangle$ and $|-\rangle$ defined in
Sec.~\ref{subsec:identical}.  If the dots are non-identical, there is
no perfect correspondence between these two pairs of states but, for
non-zero coupling, still one of them is brighter and the other is
darker. As we have shown previously \cite{karwat11}, the
phonon-induced redistribution of single exciton occupations among
these two states strongly affects spontaneous emission in a
temperature-dependent way. 
For the system of vertically stacked dots
analyzed in this paper, the amplitudes of both tunneling and F{\"o}rster  
couplings are negative, so that $\phi<0$. In this case, the darker
eigenstate has a  
higher energy and will be affected by relaxation even at low
temperatures. As we show below, phonon dynamics indeed
breaks the relative stability of the darker state and leads to
considerably accelerated decay of the excitonic occupation,  except for special
parameter choices.

While, in general, non-Markovian effects may be
important in the carrier-phonon dynamics that affects spontaneous
emission \cite{karwat11}, the phonon-related effects to be discussed
below can be understood within a Markovian picture of transitions
between energy eigenstates. The corresponding rate for a transition
between the eigenstates $|\Psi_{i}\rangle$ and $|\Psi_{j}\rangle$ is
$\gamma_{i\to j}=2\pi \sin^{2}\theta 
J(|\omega_{ij}|)|n(\omega_{ij})+1|$, 
where $i,j=1,2$, $\omega_{ij}=(E_{j}-E_{i})/\hbar$ 
and the spectral density is defined as
\begin{displaymath}
J(\omega)=\sum_{\bm{q}}\left| f_{\bm{q}} \right|^{2}
\sin^{2}\frac{k_{z}D}{2}\delta(\omega-\omega_{\bm{q}}).
\end{displaymath} 
The spectral density for the inter-dot spacing $D=8$~nm is
plotted in Fig.~\ref{fig:spdens}. Apart from the usual cutoff at 
frequencies larger than $\overline{l}/c$ due to the restriction on
the momentum non-conservation $\Delta(\hbar q)\lesssim
\hbar/\overline{l}$, where $\overline{l}$ is the 
average dot size, it is characterized by the oscillations with the
period $\Delta\omega=2\pi c/D$ which result from the double dot
structure of the system \cite{karwat11}.

\begin{figure}[t]
\centering
\includegraphics[scale=0.7]{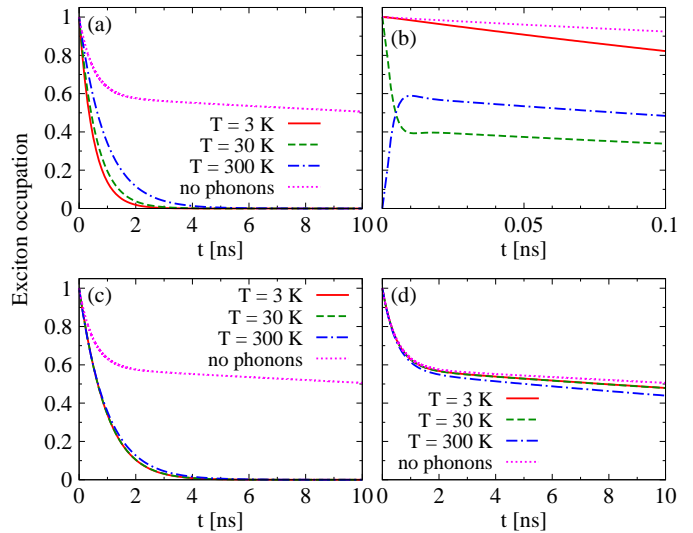}
\caption{Phonon impact on vacuum-induced coherence in a pair of QDs
with a fixed $V/\Delta$ ratio (corresponding to $\phi=-1.4$): the
comparison of the total exciton occupation in a system coupled to
phonons and without phonons. (a) $\Omega=4.0$~meV; (b) short-time
section of (a) with solid (red) line representing the total exciton
occupation, dotted (magenta) line corresponding to the total
occupation without phonons and the dashed (green) and dash-dotted
(blue) lines showing the occupations of the two dots; (c)
$\Omega=8.07$~meV (corresponding to the 3rd minimum of the spectral 
density); (d) $\Omega=13.36$~meV (5th minimum of the spectral density).}
\label{fig:VIC_ph}
\end{figure}

The phonon-related effects on the spontaneous buildup of coherence is
shown in Fig.~\ref{fig:VIC_ph}. Here we keep the mixing angle $\phi$
constant (there is, a constant ratio $V/\Delta$) and use the energy splitting
$\hbar\Omega=E_{1}-E_{2}=2\sqrt{\Delta^{2}+V^{2}}$ as a parameter.
In Fig.~\ref{fig:VIC_ph}(a) we plot the evolution of the total exciton
occupation for $\hbar\Omega=4.0$~meV, where the spectral density is
relatively large, corresponding to phonon transition rates on the
order of 1~ps$^{-1}$. One can see here that indeed the phonon-induced
relaxation suppresses the vacuum-induced coherence, which is
manifested by the rapid decay of the excitonic occupation. The role of
phonons in this strong change of the occupation dynamics is clear in
Fig.~\ref{fig:VIC_ph}(b), where we show the first 100~ps of the same
evolution. A very fast phonon-induced redistribution of occupations of
the two dots (shown in green dashed and blue dash-dotted lines) takes
place on a picosecond time scale after which the 
radiative effects are completely dominated by the phonon-induced
thermalization which, at low temperatures, forces the system to stay
in the bright state $|\Psi_{2}\rangle$. The detrimental effect of
carrier-phonon coupling can only be avoided if the phonon-induced
occupation dynamics is made slow compared to the spontaneous
emission. This can be achieved by using the oscillating form of the
phonon spectral density (see Fig.~\ref{fig:spdens}). As can be seen in
Fig.~\ref{fig:VIC_ph}(c,d),  If
choosing the energy splitting $\hbar\Omega$ such that it corresponds
to the 3rd minimum of the spectral density has little positive effect
but for $\hbar\Omega$ in the 5th minimum, the phonon effects become
very weak and the long living tail of the excitonic occupation is
restored. 

An interesting additional effect can be seen if one compares
Figs.~\ref{fig:VIC_ph}(a) and \ref{fig:VIC_ph}(d): If the phonon
effects are strong then increasing the temperature slows down the
occupation decay. However, for weak phonon influence the temperature
dependence is opposite. This can be explained by noting that in the
presence of fast phonon-induced redistribution of single exciton
occupations, the occupation of these two states remains in
quasi-equilibrium, which means that the occupation of the higher energy,
darker state increases as the temperature grows. On the contrary, if
the phonon-induced dynamics is slow the system state is almost
unperturbed and close to the spontaneously formed dark
superposition. Phonons lead to transitions out of this stable state
with intensity growing with temperature.
 
\section{Conclusions}
\label{sec:concl}

We have studied the formation of vacuum-induced coherence and the
associated long-living trapped excitonic population in a pair of
vertically stacked semiconductor QDs. We focused on the 
features that distinguish QD systems from natural 
atoms: the mismatch of transition energies, coupling,
possibly non-identical dipole moments for the optical transition, and
strong interactions with the phonon environment.

We have shown that 
the VIC effect is very sensitive to the inhomogeneity of the
QDs. Already for the fundamental transition energy mismatch 
on the order of the emission line width, the exciton occupation is
quenched. However, the destructive effect 
of the energy inhomogeneity can be strongly reduced 
by coupling between the dots. While for dots with identical magnitudes
of the interband dipole moments full stabilization of the
spontaneously formed coherence can only be achieved in the limit of
infinite coupling, in pairs of QDs with different interband dipoles it
is possible to adjust the energy 
mismatch, the coupling between the QDs, and the dipole moments in such
a way that a 
perfectly stable state forms in the spontaneous emission process and a
fraction of the initial excitonic population attains a formally infinite life
time. The non-parallel orientation of the interband dipoles, that may
be caused by heavy-light hole mixing leads to negligibly weak effects
in realistic structures. While carrier-phonon coupling typically
destroys the vacuum induced coherence on picosecond time scales, it
can be overcome by appropriately selecting the energy splitting
between the single exciton states.

These results show that the VIC effect  
can be observed in realistic systems with energy splitting on the order of
milli-electron-Volts provided that the system parameters (interband
dipoles, coupling and energy mismatch) can be controlled with
sufficient flexibility. This seems to be possible by appropriately
designing the system on the manufacturing stage and then employing the
dependence of various parameters on external fields (e.g, via Stark
effect or modification of electron-hole wave function overlap). 

Let us note, finally, that the major experimentally detectable
consequence of the appearance of vacuum-induced coherence in the
double dot system is the long-living tail in 
the exciton occupation (or in luminescence intensity). Since this
effect is of considerable amplitude and evolves on long, nanosecond
time scales it should be relatively easily detectable with
time-resolved luminescence or pump probe spectroscopy.

\begin{acknowledgments}
A. S. acknowledges support within a ``M{\l}oda Kadra 2015 Plus'' 
project, co-financed by the Polish Ministry of Science and Higher
Education (MNiSW) and the European Union within the 
European Social Fund, and within a scholarship for outstanding young
scientists granted by the Polish MNiSW. 
\end{acknowledgments}

\appendix*
\section{Interband dipole moments and
subband mixing}
\label{sec:app}

In this appendix, we discuss the effect of light hole admixture on the
non-parallel alignment of the interband dipoles for (nominally) 
heavy hole transitions. 

Within the 4-band Luttinger model \cite{luttinger56},
the electron and hole wave functions are
\begin{eqnarray*}
\Psi^{(\alpha)}_{\mathrm{h}}\!(\bm{r},s) & = & 
\sum_{\lambda}u_{\lambda}
\!(\bm{r},s)\varphi^{(\alpha)}_{\mathrm{h}\lambda}\!(\bm{r}) \\
\Psi^{(\alpha)}_{\mathrm{e}}\!(\bm{r},s) & = & 
u_{\mathrm{c}\frac{1}{2}}\!(\bm{r},s)\varphi^{(\alpha)}_{\mathrm{e}}\!(\bm{r}),
\end{eqnarray*}
where $\alpha=1,2$ refers to a higher or lower QD, respectively,
$u_{\lambda}$ is a Bloch function for the valence  
subband $\lambda$ (heavy holes: $\lambda=\pm 3/2$, light holes: $\lambda=\pm 1/2$), $\bm{r}$ is the space coordinate, 
$s$ denotes spin, $u_{\mathrm{c}\frac{1}{2}}$ is an electron Bloch
function with a fixed spin $s=1/2$ (for definiteness) and  
$\varphi^{(\alpha)}_{\mathrm{e/h}}$ refer to electron and hole envelope functions, respectively.

The matrix element of the interband dipole moment for the $\alpha$th QD is
\begin{displaymath}
\bm{d}_{\alpha}=\sum_{s}\!\int\! d^{3}r\Psi^{(\alpha)}_{\mathrm{h}}
\!(\bm{r},s)e\bm{r}\Psi^{(\alpha)}_{\mathrm{e}}\!(\bm{r},s).
\end{displaymath}     
To calculate the above integral we sum over unit cells (labeled by
$\bm{R}$) and integrate over one unit cell ($\bm{\xi}$ labels  
the position within the cell). The envelope functions vary slowly
which allows us to assume that they are constant over one  
unit cell,
$\varphi_{\mathrm{e/h}}(\bm{R}+\bm{\xi})\approx\varphi_{\mathrm{e/h}}(\bm{R})$. The
Bloch functions are periodic,  
$u(\bm{R}+\bm{\xi},s)=u(\bm{R},s)$, and orthogonal for different
bands. As a result, we can write \cite{haug04}
\begin{eqnarray*}
\bm{d}_{\alpha} & = & 
\frac{1}{v}\sum_{s,\lambda}\int d^{3}
\mathrm{R}\int\limits_{u.c.}\!d^{3}\xi \\
& & \times u_{\lambda}\!(\bm{\xi},s)
\varphi^{(\alpha)}_{\mathrm{h}\lambda}\!(\bm{R})
e\left(\bm{R}+\bm{\xi}\right)u_{\mathrm{c}\frac{1}{2}}\!(\bm{\xi},s)
\varphi^{(\alpha)}_{\mathrm{e}}\!(\bm{R}) \\
& = & \sum\limits_{\lambda}a^{(\alpha)}_{\lambda}\bm{d}_{\lambda,\frac{1}{2}}.
\end{eqnarray*}
Here $\bm{d}_{\lambda,\frac{1}{2}}$ is the bulk interband dipole moment,
\begin{displaymath}
\bm{d}_{\lambda,\frac{1}{2}} = 
\frac{1}{v}\sum_{s}\int\limits_{u.c.}\!d^{3}\xi
u_{\lambda}\!(\bm{\xi},s)
e\bm{\xi}u_{c\frac{1}{2}}\!(\bm{\xi},s),
\end{displaymath}  
$v$ is the unit cell volume,
\begin{displaymath}
a^{(\alpha)}_{\lambda} = \int d^{3}R \varphi^{(\alpha)}_{\mathrm{h}\lambda}\!(\bm{R})\varphi^{(\alpha)}_{\mathrm{e}}\!(\bm{R})
\end{displaymath} 
is the envelop function overlap integral,
and we have replaced the summation over unit cells with
integration over $\bm{R}$.

We are investigating bright heavy hole excitons, hence
$a^{(\alpha)}_{-3/2}\sim 1$ and the other coefficients
$a^{(\alpha)}_{\lambda}$ 
are much smaller. 
The non-vanishing bulk dipole moment matrix elements involving the
spin-$1/2$ electron state are
\cite{haug04,axt94b} 
\begin{eqnarray*}
\bm{d}_{-\frac{3}{2},\frac{1}{2}} = 
-\sqrt{3} \bm{d}_{-\frac{1}{2},\frac{1}{2}}^{*}
&=&\frac{1}{\sqrt{2}}d_{0}\left(
\begin{array}{c}
-1 \\ i \\ 0
\end{array}
\right);
\\
\bm{d}_{\frac{1}{2},\frac{1}{2}}&=&\sqrt{\frac{2}{3}}d_{0}\left(
\begin{array}{c}
0 \\ 0 \\ 1
\end{array}
\right).
\end{eqnarray*}

Because the wave functions differ for the two QDs, the values of the
heavy hole overlap integrals $a^{(\alpha)}_{-\frac{3}{2}}$  
may vary. Since, to the leading order,
\begin{eqnarray*}
|\bm{d}_{\alpha}| & = & 
|d_{0}|\left(\left|a^{(\alpha)}_{-\frac{3}{2}}\right|^{2}
+\frac{2}{3}\left|a^{(\alpha)}_{-\frac{1}{2}}\right|^{2}
+\frac{1}{3}\left|a^{(\alpha)}_{\frac{1}{2}}\right|^{2}\right)^{\frac{1}{2}}\\
& \approx & \left|a^{(\alpha)}_{-\frac{3}{2}}d_{0} \right|
\left(1+\frac{2\left|a^{(\alpha)}_{-\frac{1}{2}}\right|^{2}
+\left|a^{(\alpha)}_{\frac{1}{2}}\right|^{2}}{
6\left|a^{(\alpha)}_{-\frac{3}{2}}\right|^{2}}
\right),
\end{eqnarray*} 
the difference of the overlap integrals leads to different magnitudes
of the dipole moments $|\bm{d}_{\alpha}|$, $\alpha = 1,2$.  
Moreover, one has
\begin{displaymath}
\bm{d}_{1}\cdot\bm{d}^{*}_{2}=
a^{(1)}_{-\frac{3}{2}}a^{(2)*}_{-\frac{3}{2}}|d_{0}|^{2}
\left(1+\frac{2}{3} 
\frac{a^{(1)}_{-\frac{1}{2}} a^{(2)*}_{-\frac{1}{2}}}{
a^{(1)}_{-\frac{3}{2}} a^{(2)*}_{-\frac{3}{2}}}
+\frac{1}{3} \frac{a^{(1)}_{\frac{1}{2}} a^{(2)*}_{\frac{1}{2}}}{
a^{(1)}_{-\frac{3}{2}} a^{(2)*}_{-\frac{3}{2}}} \right).
\end{displaymath}
Hence, 
\begin{eqnarray}
\lefteqn{\left| \hat{\bm{d}}_{1}\cdot\hat{\bm{d}}^{*}_{2} \right|
 =  \frac{|\bm{d}_{1}\cdot\bm{d}^{*}_{2}|}{|\bm{d}_{1}||\bm{d}_{2}|}}
 \nonumber \\
&& \approx  1-
\frac{1}{3}
\left|\frac{a^{(1)}_{-\frac{1}{2}}}{a^{(1)}_{-\frac{3}{2}}}
-\frac{a^{(2)}_{-\frac{1}{2}}}{a^{(2)}_{-\frac{3}{2}}}\right|^{2}
-\frac{1}{6}
\left|\frac{a^{(1)}_{\frac{1}{2}}}{a^{(1)}_{-\frac{3}{2}}}
-\frac{a^{(2)}_{\frac{1}{2}}}{a^{(2)}_{-\frac{3}{2}}}
\right|^{2}.
\label{dipole1}
\end{eqnarray}
Thus, if the light hole 
admixture is different for the two dots,
\begin{displaymath}
\frac{a^{(1)}_{\pm\frac{1}{2}}}{a^{(1)}_{-\frac{3}{2}}} 
\neq \frac{a^{(2)}_{\pm\frac{1}{2}}}{a^{(2)}_{-\frac{3}{2}}} 
\end{displaymath}
then $\hat{\bm{d}}_{1}\cdot\hat{\bm{d}}^{*}_{2}\neq 1$,
 that is, the dipoles are non-parallel.

As the subband mixing is typically small in self-assembled structures
the angle $\theta$ between the dipole moments is small.  
Therefore, one can write
\begin{equation}
\label{dipole2}
\hat{\bm{d}}_{1}\cdot\hat{\bm{d}}^{*}_{2}
=e^{i\eta} \cos\theta
\approx e^{i\eta}\left(  1-\frac{1}{2}\theta^{2}  \right),
\end{equation}
where $\eta$ is an irrelevant phase.
Comparing Eqs. (\ref{dipole1}) and (\ref{dipole2}) one gets
\begin{equation*}
\theta=\left(\frac{2}{3}
\left|\frac{a^{(1)}_{-\frac{1}{2}}}{a^{(1)}_{-\frac{3}{2}}}
-\frac{a^{(2)}_{-\frac{1}{2}}}{a^{(2)}_{-\frac{3}{2}}}\right|^{2}
+\frac{1}{3}
\left|\frac{a^{(1)}_{\frac{1}{2}}}{a^{(1)}_{-\frac{3}{2}}}
-\frac{a^{(2)}_{\frac{1}{2}} }{a^{(2)}_{-\frac{3}{2}}}
\right|^{2}
\right)^{\frac{1}{2}}.
\end{equation*}
%
%


\end{document}